\documentclass[runningheads]{llncs}
\usepackage{graphicx}
\usepackage{amssymb, amsmath}
\DeclareMathOperator*{\argmin}{arg\,min}
\begin{document}
\title{Physics-informed self-supervised deep learning reconstruction for accelerated first-pass perfusion cardiac MRI}
\titlerunning{Physics-informed self-supervised DL reconstruction for FPP-CMR}
\author{Elena Mart{\'i}n-Gonz{\'a}lez\inst{1}\orcidID{0000-0002-5922-4960} \and Ebraham Alskaf\inst{2}\orcidID{0000-0003-4007-6854} \and Amedeo Chiribiri\inst{2}\orcidID{0000-0003-3394-4289} \and Pablo Casaseca-de-la-Higuera\inst{1}\orcidID{0000-0003-1565-0842} \and Carlos Alberola-López\inst{1}\orcidID{0000-0003-3684-0055} \and Rita G. Nunes\inst{3}\orcidID{0000-0001-7425-5717} \and Teresa Correia\inst{2,4}\orcidID{0000-0002-1606-9550}}
\authorrunning{E. Mart{\'i}n-Gonz{\'a}lez et al.}
\institute{Laboratorio de Procesado de Imagen, ETSI Telecomunicaci{\'o}n, Universidad de Valladolid, Valladolid, Spain \\ \email{emargon@lpi.tel.uva.es}\and
School of Biomedical Engineering and Imaging Sciences, King’s College London, London, United Kingdom\\ \and
Institute for Systems and Robotics and Department of Bioengineering, Instituto Superior Técnico, Universidade de Lisboa, Lisbon, Portugal \and Centre for Marine Sciences - CCMAR, Faro, Portugal}
\maketitle              % typeset the header of the contribution
\begin{abstract}
%The abstract should briefly summarize the contents of the paper in 15--250 words.
First-pass perfusion cardiac magnetic resonance (FPP-CMR) is becoming an essential non-invasive imaging method for detecting deficits of myocardial blood flow, allowing the assessment of coronary heart disease. Nevertheless, acquisitions suffer from relatively low spatial resolution and limited heart coverage. Compressed sensing (CS) methods have been proposed to accelerate FPP-CMR and achieve higher spatial resolution. However, the long reconstruction times have limited the widespread clinical use of CS in FPP-CMR. Deep learning techniques based on supervised learning have emerged as alternatives for speeding up reconstructions. However, these approaches require fully sampled data for training, which is not possible to obtain, particularly high-resolution FPP-CMR images. Here, we propose a physics-informed self-supervised deep learning FPP-CMR reconstruction approach for accelerating FPP-CMR scans and hence facilitate high spatial resolution imaging. The proposed method provides high-quality FPP-CMR images from 10x undersampled data without using fully-sampled reference data.

\end{abstract}
\section{Introduction}
Coronary artery disease (CAD) is the occlusion of the coronary arteries usually caused by atherosclerosis, which causes abnormalities in blood flow to the heart. Non-invasive imaging techniques that are widely used clinically for the evaluation of CAD are single photon emission computerized tomography (SPECT) and positron emission tomography (PET), but the reference for non-invasive myocardial perfusion quantification is PET~\cite{heo2014noninvasive}. However, the clinical value of first-pass perfusion cardiac magnetic resonance (FPP-CMR) has been shown in comparison to these techniques \cite{foley2018cardiovascular,hendel2016cmr,heo2014noninvasive,schwitter2013mr}, having emerged as an alternative way of detecting blood flow anomalies without the use of potentially harmful ionising radiation. In addition, FPP-CMR has other advantages, such as higher spatial resolution, wider availability and lower scan cost compared to PET. 

FPP-CMR time frames must be acquired in real-time to capture the rapid passage of a contrast agent bolus through the heart, and hence, the spatial resolution and coverage of the heart is compromised. Thus, undersampled reconstruction methods have been proposed to accelerate FPP-CMR acquisitions as a means to improve spatial resolution and heart coverage \cite{lingala2011accelerated,otazo2010combination,vitanis2011high}. However, these methods can lead to long reconstruction times. In this work, we aim to speed up reconstructions and obtain the  contrast-enhanced dynamic image series from undersampled FPP-CMR using deep learning (DL). Then, these images will be used to generate quantitative perfusion maps using a tracer kinetic model \cite{correia2019model,hsu2018diagnostic,kellman2017myocardial}. DL techniques have already been used in magnetic resonance image (MRI) reconstruction. Work has been reported on knee \cite{aggarwal2018modl,kocanaogullari2019deep,yaman2020self}, brain \cite{aggarwal2018modl,do2020reconstruction,kocanaogullari2019deep,yaman2020self} and cardiac \cite{huang2019mri,schlemper2017deep} MRI, using both supervised \cite{aggarwal2018modl,do2020reconstruction,kocanaogullari2019deep} and self-supervised learning \cite{liu2021magnetic,yaman2020self}. Occasionally, the network is unrolled to mimic a compressed sensing (CS) iterative reconstruction problem, giving rise to a cascade of convolutional neural networks (CNNs) \cite{aggarwal2018modl,kocanaogullari2019deep,schlemper2017deep}. The problem with supervised learning techniques is the need to have fully sampled reference images to train the network, which are not available in FPP-CMR, particularly at high spatial resolutions.

Even though the field of MRI reconstruction with DL is currently an active area, to our knowledge, self-supervised DL techniques have not been applied to FPP-CMR reconstruction. In this work, a SElf-Supervised aCcelerated REconsTruction (SECRET) DL framework for FPP-CMR is proposed to directly reconstruct contrast-enhanced dynamic image series from undersampled (k,t)-space data. 
\section{Methods}
For completeness, a conventional FPP-CMR CS reconstruction will be described. We will also describe our proposed method,  SECRET, as well as the  Model Based Deep Learning Architecture for Inverse Problems (MoDL) \cite{aggarwal2018modl}, which will be used for comparison.
\subsection{Conventional FPP-CMR reconstruction}

CS methods can be used to reconstruct dynamic images from undersampled data. For example, FPP-CMR images $\mathbf{s}$ can be obtained from undersampled data $\mathbf{d}_u$ using CS by solving the following optimisation problem:

\begin{equation}
\begin{split}
    \mathbf{\hat{s}} = \argmin_{\mathbf{s}} \{\Vert \mathbf{d}_u - \mathbf{E}\mathbf{s}\Vert_2^2 + \lambda_1 \Vert \nabla_s \mathbf{s} \Vert_1 + \lambda_2 \Vert \nabla_t \mathbf{s}\Vert_1\}
\end{split}
\end{equation}

\noindent where $\text{\textbf{E}}=\text{\textbf{A}}\mathcal{F}$, $\mathbf{A}$ is the (k,t)-space sampling trajectory, $\mathcal{F}$ is the Fourier transform, $\lambda_1$ and $\lambda_2$ are regularization parameters and $\nabla_s$ and $\nabla_t$ are the finite differences operators along the spatial and temporal dimensions, respectively.

\subsection{Supervised learning reconstruction: MoDL}

MoDL combines the power of DL with model-based approaches \cite{aggarwal2018modl}. It uses a CNN as a denoiser and applies it as a regulariser to solve the optimisation problem given by:
\begin{equation}
    \mathbf{s}_{k+1} = \argmin_{\mathbf{s}} \Vert \mathbf{d}_u - \mathbf{E}\mathbf{s}\Vert_2^2 + \lambda \Vert\mathbf{s} - \mathbf{z}_k\Vert_2^2
\end{equation}
\begin{equation}
    \mathbf{s}_{k+1} = \left( \text{\textbf{E}}^H \text{\textbf{E}} + \lambda{\mathcal{I}} \right)^{-1} \left( \text{\textbf{E}}^H\mathbf{d_u} + \lambda\mathbf{z}_k \right)
\end{equation}
where $k$ denotes the $k$-th iteration and $\mathbf{z}_k$ is the denoised version of $\mathbf{s}_k$, obtained through a CNN network. MoDL requires supervised learning to optimise the denoiser network. The data consistency layer is immediate by conjugate gradient blocks, but as the input is $\mathbf{z}_k$ and the output is $\mathbf{s}_{k+1}$, which, in turn, generates a $\mathbf{z}_{k+1}$, this requires iterating until convergence. The iterative algorithm is unrolled for a fixed number of iterations, K, in which the weights or parameters to be optimised are shared. 

The MoDL method has the zero-filled reconstruction, the coil sensitivities and the subsampling mask as inputs, but it also needs the fully sampled images ---which are hardly available for the case of FPP-CMR at high spatial resolution--- for training. The loss is defined as the mean square error between ${s}_K$ and the desired image $t$: $C =\sum\limits_{i=1}^{Nsamples} \Vert \mathbf{s}_K(i) - \mathbf{t}(i) \Vert^2
$, where $\mathbf{t}(i)$ is the $i$-th target image.

\subsection{SECRET reconstruction}
\begin{figure}[htb!]
\centering
\includegraphics[width=\textwidth]{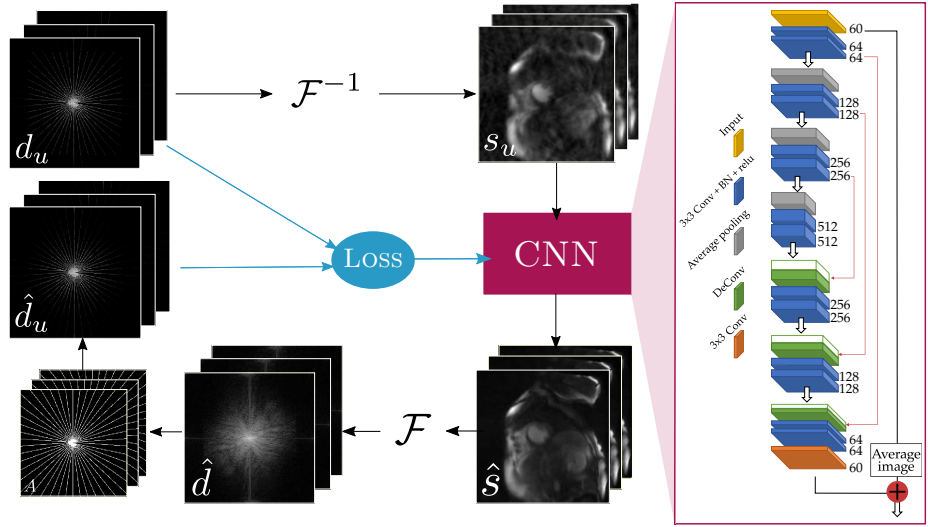}
\caption{Flow chart illustrating the proposed SECRET method for FPP-CMR. Blue lines represent steps that only take place during training. The inputs of the framework are the undersampled (k,t)-space data $\mathbf{d}_u$ and the (k,t)-sampling masks $A$, resulting in the reconstructed contrast-enhanced dynamic images $\mathbf{\hat{s}}$ as output, and $\mathbf{\hat{d}}_u$ if required.}
\label{fig:diagram}
\end{figure}

The proposed SECRET method directly reconstructs contrast-enhanced dynamic images from the undersampled (k,t)-space data. Considering only the undersampled (k,t)-space data when enforcing data consistency, we can train networks without the need for fully sampled images, simply by making use of the physical models in the reconstruction \cite{liu2021magnetic}. This framework can be formulated as follows:
\begin{equation}
    \hat{\theta} = \argmin_{\theta} \Vert \mathbf{d}_u - \mathbf{A}\mathcal{F}\text{C}(\mathbf{s}_u|\boldsymbol{\theta})\Vert_2 
\end{equation}
where $\text{C}(\mathbf{s}_u|\boldsymbol{\theta})$ is the output of a CNN, with $\boldsymbol{\theta}$ the parameter vector to be optimised. Figure~\ref{fig:diagram} shows the steps necessary for training our proposed SECRET method for FPP-CMR. First, undersampled (k,t)-space data $\mathbf{d}_u$ is transformed to the image domain, obtaining $\mathbf{s}_u$. Then, $\mathbf{s}_u$ enters the CNN to provide the reconstructed contrast-enhanced dynamic images $\mathbf{\hat{s}}$. These images are then transformed back to (k,t)-space $\mathbf{\hat{d}}$ and subsampling masks are applied, thus obtaining the undersampled version $\mathbf{\hat{d}}_u$. Finally, the loss is computed with $\mathbf{\hat{d}}_u$ and the input $\mathbf{d}_u$, to guide the training phase.

The CNN is based on the well-known U-Net~\cite{unet}, widely used in medical imaging. Skip connections are included to maintain information from previous layers, as well as to avoid the problem of vanishing gradients during backpropagation. At the end of the CNN, residual learning has been appended as in \cite{liu2021magnetic}, adding the average image of the input $\mathbf{s}_u$.

%
%\subsection{TK parameters maps}
%
%Patlak model \cite{patlak1983graphical}
%
\subsection{Dataset}
Rest and stress FPP-CMR acquisitions were performed in 21 patients using a single-bolus injection of $0.05$ mmol/kg Gadobutrol (Gadovist; Bayer, Germany) and a 1.5T CMR scanner (MAGNETOM Aera, Siemens Healthineers, Erlangen, Germany) with an 18-channel chest-coil and a 32-channel spine coil. A free-breathing FLASH perfusion dual-sequence \cite{kellman2017myocardial} was used to acquire a low-resolution image with low T1-sensitivity for estimating the arterial input function and three short-axis slices (basal, mid and apical) for high resolution myocardial perfusion imaging using the following parameters: $\text{FOV}=340 \times 308$ mm$^2$, in-plane resolution = $2.2 \times 2.2$ mm$^2$, $\text{slice thickness}=10 \text{mm}$, TR/TE = 2.1/1ms, flip angle = $8^{\circ}$, parallel imaging acceleration factor 3, saturation recovery time = 100 ms, total scan duration = 60s, $\text{contrast agent relaxivity}=5.0 \text{L/mmol s}$. Undersampled datasets were generated for $3\times$, $6\times$ and $10\times$ acceleration factors, following a radial (k,t)-sampling trajectory.

\noindent \textbf{Preprocessing}{ A first step to ensure that all data had the same size, both spatially and temporally, prior to being fed to the CNN, consisted of resizing the DICOM images to obtain a spatial resolution of $2 \times 2$ mm$^2$, padding the k-space to obtain an image size of $256\times256$ pixels, and interpolating each slice to a fixed number of frames (60 frames). A final step included intensity normalisation so that all contrast-enhanced dynamic image series present intensities between 0 and 1, without losing the contrast variation between frames. In addition, image pre-registration was also carried out to correct for respiratory motion.}

\noindent \textbf{Image quality metrics} Image quality was assessed in terms of peak signal-to-noise ratio (PSNR), structural similarity index measure (SSIM) and normalized root mean square error (NRMSE) between the reference images and  reconstructions obtained with the SECRET, MoDL and CS (10x only) methods. 

\subsection{Implementation details}
Patients were randomly split into training, validation and test subsets (60\%, 16\% and 24\%, respectively). Each slice is fed into the SECRET framework so that the time frames are stacked in depth, creating a multi-channel image. The proposed method is implemented in Python with Tensorflow \cite{tensorflow2015-whitepaper} and Keras \cite{chollet2015keras}, and it took about half an hour of training using the Adam optimizer \cite{kingma2014adam} with a learning rate of $10^{-4}$ consuming about 3 GB of GPU memory for 100 epochs on one Intel\textsuperscript{\tiny\textregistered} Core\textsuperscript{\tiny{TM}} i7-4790 CPU @ 3.60GHz with 16 GB RAM and one NVIDIA GeForce RTX 2080 Ti GPU. The MoDL training for K=1 and 100 epochs took one hour and a half and the MoDL training for K=10 and 200 epochs took forty-five hours using the same hardware. Note that after training the SECRET method, it provides a reconstruction of a complete contrast-enhanced dynamic image series in less than a second.

\section{Results and discussion}

\begin{figure}[htb!]
\centering
\includegraphics[width=0.75\textwidth]{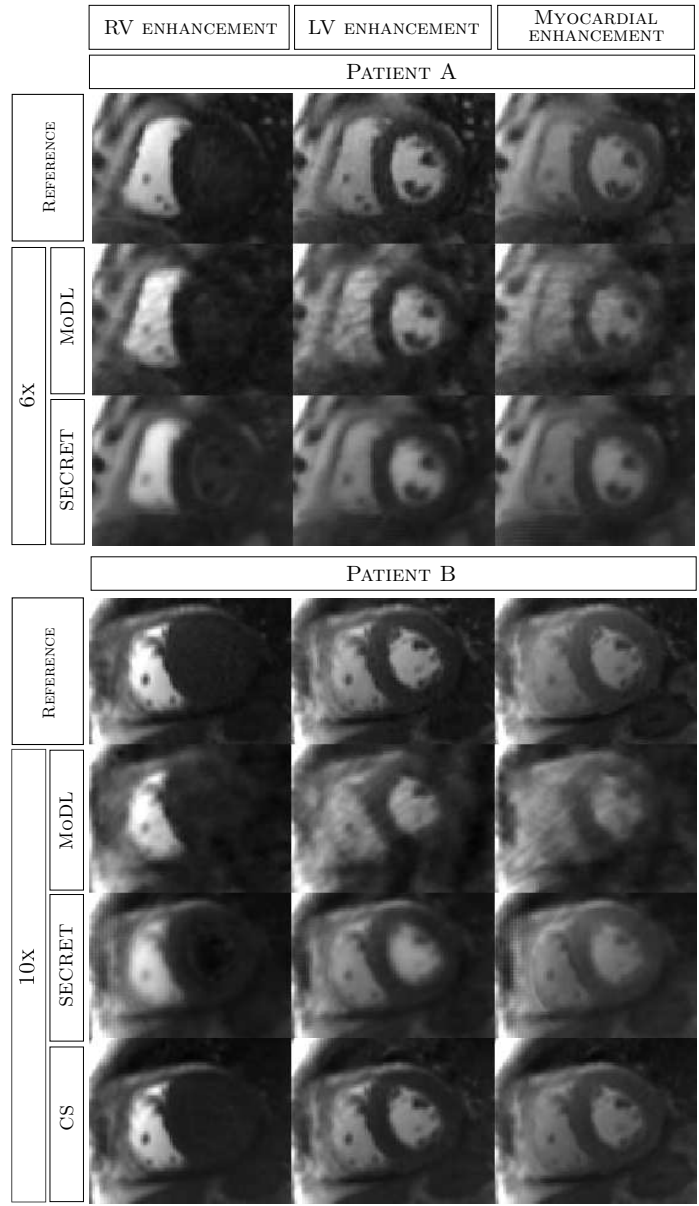}
\caption{SECRET and MoDL (K=1) reconstructions obtained from $6\times$ and $10\times$ undersampled FPP-CMR data for two representative subjects. The reference images are displayed for comparison, in addition to CS reconstruction for $10\times$. The right ventricle (RV), left ventricle (LV) and myocardial enhancement time frames are shown for one short axis slice. } \label{fig:recons}
\end{figure}

Figure~\ref{fig:recons} shows the SECRET reconstructions obtained for two representative patients from $6\times$ and $10\times$ undersampled (k,t)-space data together with the reference and MoDL (K=1) reconstructions. CS reconstruction is also shown for $10\times$. Three different time frames are shown, corresponding to right ventricle (RV), left ventricle (LV) and myocardial enhancement. Although the SECRET reconstructions are slightly blurred, due to residual learning from the average image of the CNN input (which is blurred due to residual motion), it can be seen that they have better quality than the images obtained with MoDL trained in the same amount of time. Moreover, SECRET images maintain the variability of contrast that exists between frames in addition to not losing the structure of the heart.

\begin{figure}[t]
\centering
\includegraphics[width=0.9\textwidth]{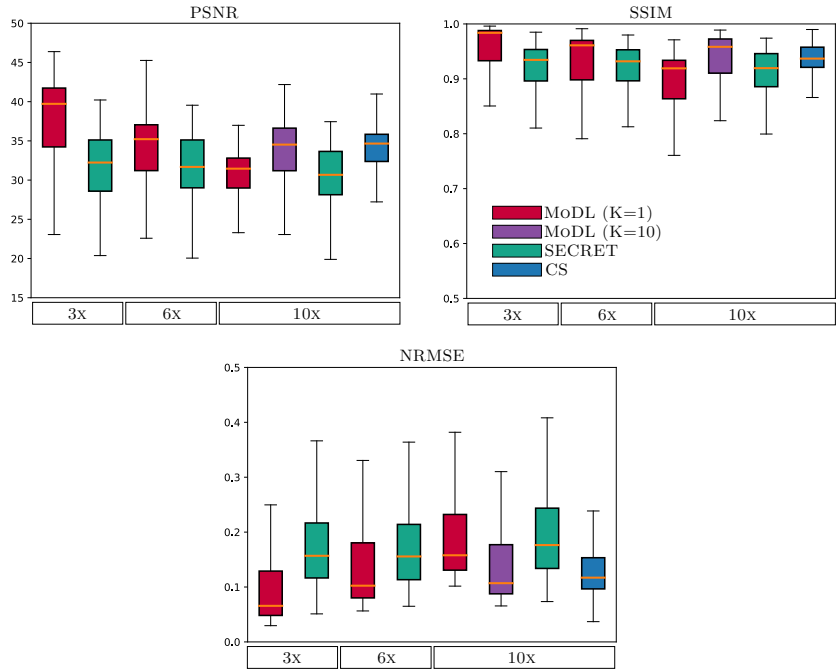}
\caption{PSNR, SSIM and NRMSE between the reference images and the reconstructions obtained with SECRET and MoDL methods, for $3\times$, $6\times$ and $10\times$ acceleration factors, for all patients in the test dataset.}
\label{fig:metrics}
\end{figure}

Figure~\ref{fig:metrics} shows results of the FPP-CMR reconstructions in terms of PSNR, SSIM and NRMSE. While the performance of MoDL becomes noticeably worse as the acceleration rate increases, SECRET maintains good image quality even at high acceleration rates. For the 10x accelerated reconstructions, the \textbf{median} (interquartile range): PSNR was \textbf{34.66} (3.47), \textbf{31.46} (3.81), \textbf{34.52} (5.43), \textbf{30.67} (5.52); SSIM was \textbf{0.94} (0.04), \textbf{0.92} (0.07), \textbf{0.96} (0.06), \textbf{0.92} (0.06); NRMSE was \textbf{0.12} (0.06), \textbf{0.16} (0.10), \textbf{0.11} (0.09), \textbf{0.17} (0.11) for CS, MoDL (K=1), MoDL (K=10) and SECRET methods, respectively. The image quality metrics indicate that SECRET images maintain a more stable agreement with the reference as the acceleration factor is increased than MoDL images, which deteriorate with higher acceleration. CS and MoDL (K=10) show the best agreement with the reference, but reconstructions take $\sim$87.08s and $\sim$1.99s, respectively, whereas MoDL (K=1) takes $\sim$0.21s and SECRET only 0.15s.

\begin{figure}[t]
\centering
\includegraphics[width=0.75\textwidth]{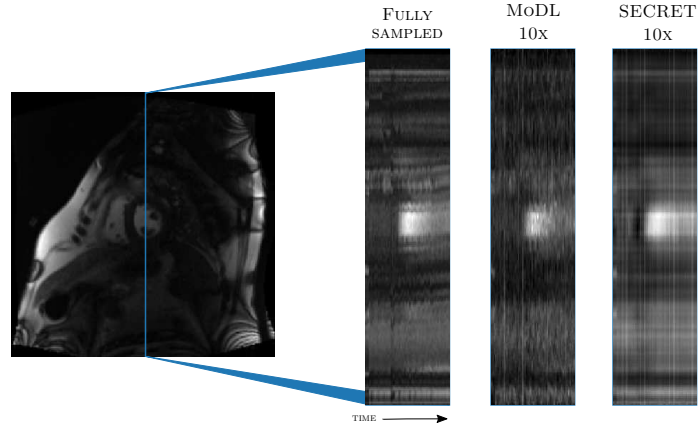}
\caption{Representative image profile across the heart demonstrating that the SECRET framework improves consistency across time frames.} \label{fig:motion}
\end{figure}

Figure~\ref{fig:motion} shows a 1D projection of the dynamic images through time, for a given slice. Note that although the images have been pre-registered, there is still some residual motion. SECRET does not include any explicit regularisation term, however, due to the residual learning performed by the network all reconstructions provided by the framework are inherently corrected. Such good PSNR, SSIM and NRMSE values obtained when the reference images are affected by little respiratory motion, would certainly improve if some regularisation were added. This would enable even higher acceleration rates. Regularisation schemes will thus be investigated in a future study.

\begin{figure}[htb!]
\centering
\includegraphics[width=0.8\textwidth]{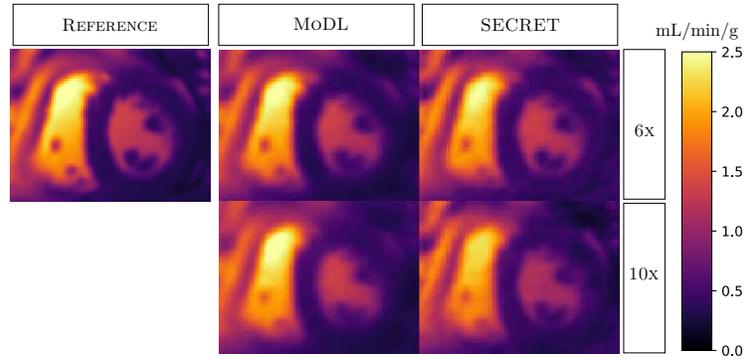}
\caption{Quantitative maps ($K^{\text{Trans}}$) obtained from $6\times$ and $10\times$ undersampled data using MoDL and the SECRET methods. The reference image is displayed for comparison.} \label{fig:ktrans}
\end{figure}

Quantitative parameter maps were estimated from the FPP-CMR reconstructions, showing the potential of the technique for an objective and operator-independent analysis of myocardial perfusion. Figure~\ref{fig:ktrans} displays the contrast transfer coefficient ($K^{\text{Trans}}$) map estimated from fully sampled, $6\times$ and $10\times$ undersampled patient data using the MoDL and SECRET methods, through the Patlak model \cite{patlak1983graphical}. The image quality of the quantitative maps obtained from the SECRET reconstruction at accelerations $6\times$ and $10\times$ is comparable to the reference images, showing less blurring than MoDL maps.

\section{Conclusion}
A physics-informed self-supervised deep learning reconstruction framework for accelerating FPP-CMR scans has been described. The proposed SECRET method provides FPP-CMR reconstructions directly from the undersampled (k,t)-space data and does not require fully sampled reference data. Compared with state-of-the-art approaches, the SECRET method maintains good quality reconstructions for higher acceleration rates, with low training times and very fast reconstruction times. The proposed SECRET method shows promising results, with the potential for improvement coupled with explicit regularization, which will be explored in future work.

\bibliographystyle{splncs04}
\bibliography{bibliography}

\begin{thebibliography}{10}
\providecommand{\url}[1]{\texttt{#1}}
\providecommand{\urlprefix}{URL }
\providecommand{\doi}[1]{https://doi.org/#1}

\bibitem{tensorflow2015-whitepaper}
Abadi, M., et~al.: {TensorFlow}: Large-scale machine learning on heterogeneous
  systems (2015), \url{https://www.tensorflow.org/}, software available from
  tensorflow.org

\bibitem{aggarwal2018modl}
Aggarwal, H.K., Mani, M.P., Jacob, M.: Modl: Model-based deep learning
  architecture for inverse problems. IEEE transactions on medical imaging
  \textbf{38}(2),  394--405 (2018)

\bibitem{chollet2015keras}
Chollet, F., et~al.: Keras. \url{https://keras.io} (2015)

\bibitem{correia2019model}
Correia, T., Schneider, T., Chiribiri, A.: Model-based reconstruction for
  highly accelerated first-pass perfusion cardiac mri. In: International
  Conference on Medical Image Computing and Computer-Assisted Intervention. pp.
  514--522. Springer (2019)

\bibitem{do2020reconstruction}
Do, W.J., Seo, S., Han, Y., Ye, J.C., Choi, S.H., Park, S.H.: Reconstruction of
  multicontrast mr images through deep learning. Medical physics
  \textbf{47}(3),  983--997 (2020)

\bibitem{foley2018cardiovascular}
Foley, J.R.J.: Cardiovascular Magnetic Resonance Imaging for the Investigation
  of Ischaemic Heart Disease. Ph.D. thesis, University of Leeds (2018)

\bibitem{hendel2016cmr}
Hendel, R.C., Friedrich, M.G., Schulz-Menger, J., Zemmrich, C., Bengel, F.,
  Berman, D.S., Camici, P.G., Flamm, S.D., Le~Guludec, D., Kim, R., et~al.: Cmr
  first-pass perfusion for suspected inducible myocardial ischemia. JACC:
  Cardiovascular imaging  \textbf{9}(11),  1338--1348 (2016)

\bibitem{heo2014noninvasive}
Heo, R., Nakazato, R., Kalra, D., Min, J.K.: Noninvasive imaging in coronary
  artery disease. In: Seminars in nuclear medicine. vol.~44, pp. 398--409.
  Elsevier (2014)

\bibitem{hsu2018diagnostic}
Hsu, L.Y., Jacobs, M., Benovoy, M., Ta, A.D., Conn, H.M., Winkler, S., Greve,
  A.M., Chen, M.Y., Shanbhag, S.M., Bandettini, W.P., et~al.: Diagnostic
  performance of fully automated pixel-wise quantitative myocardial perfusion
  imaging by cardiovascular magnetic resonance. JACC: Cardiovascular Imaging
  \textbf{11}(5),  697--707 (2018)

\bibitem{huang2019mri}
Huang, Q., Yang, D., Wu, P., Qu, H., Yi, J., Metaxas, D.: Mri reconstruction
  via cascaded channel-wise attention network. In: 2019 IEEE 16th International
  Symposium on Biomedical Imaging (ISBI 2019). pp. 1622--1626. IEEE (2019)

\bibitem{kellman2017myocardial}
Kellman, P., Hansen, M.S., Nielles-Vallespin, S., Nickander, J., Themudo, R.,
  Ugander, M., Xue, H.: Myocardial perfusion cardiovascular magnetic resonance:
  optimized dual sequence and reconstruction for quantification. Journal of
  Cardiovascular Magnetic Resonance  \textbf{19}(1),  1--14 (2017)

\bibitem{kingma2014adam}
Kingma, D.P., Ba, J.: Adam: A method for stochastic optimization. arXiv
  preprint arXiv:1412.6980  (2014)

\bibitem{kocanaogullari2019deep}
Kocanaogullari, D., Eksioglu, E.M.: Deep learning for mri reconstruction using
  a novel projection based cascaded network. In: 2019 IEEE 29th International
  Workshop on Machine Learning for Signal Processing (MLSP). pp.~1--6. IEEE
  (2019)

\bibitem{lingala2011accelerated}
Lingala, S.G., Hu, Y., DiBella, E., Jacob, M.: Accelerated dynamic mri
  exploiting sparsity and low-rank structure: kt slr. IEEE transactions on
  medical imaging  \textbf{30}(5),  1042--1054 (2011)

\bibitem{liu2021magnetic}
Liu, F., Kijowski, R., El~Fakhri, G., Feng, L.: Magnetic resonance parameter
  mapping using model-guided self-supervised deep learning. Magnetic Resonance
  in Medicine  (2021)

\bibitem{otazo2010combination}
Otazo, R., Kim, D., Axel, L., Sodickson, D.K.: Combination of compressed
  sensing and parallel imaging for highly accelerated first-pass cardiac
  perfusion mri. Magnetic resonance in medicine  \textbf{64}(3),  767--776
  (2010)

\bibitem{patlak1983graphical}
Patlak, C.S., Blasberg, R.G., Fenstermacher, J.D.: Graphical evaluation of
  blood-to-brain transfer constants from multiple-time uptake data. Journal of
  Cerebral Blood Flow \& Metabolism  \textbf{3}(1), ~1--7 (1983)

\bibitem{unet}
Ronneberger, O., Fischer, P., Brox, T.: U-net: Convolutional networks for
  biomedical image segmentation. In: International Conference on Medical image
  computing and computer-assisted intervention. pp. 234--241. Springer (2015)

\bibitem{schlemper2017deep}
Schlemper, J., Caballero, J., Hajnal, J.V., Price, A.N., Rueckert, D.: A deep
  cascade of convolutional neural networks for dynamic mr image reconstruction.
  IEEE transactions on Medical Imaging  \textbf{37}(2),  491--503 (2017)

\bibitem{schwitter2013mr}
Schwitter, J., Wacker, C.M., Wilke, N., Al-Saadi, N., Sauer, E., Huettle, K.,
  Sch{\"o}nberg, S.O., Luchner, A., Strohm, O., Ahlstrom, H., et~al.: Mr-impact
  ii: Magnetic resonance imaging for myocardial perfusion assessment in
  coronary artery disease trial: perfusion-cardiac magnetic resonance vs.
  single-photon emission computed tomography for the detection of coronary
  artery disease: a comparative multicentre, multivendor trial. European heart
  journal  \textbf{34}(10),  775--781 (2013)

\bibitem{vitanis2011high}
Vitanis, V., Manka, R., Giese, D., Pedersen, H., Plein, S., Boesiger, P.,
  Kozerke, S.: High resolution three-dimensional cardiac perfusion imaging
  using compartment-based k-t principal component analysis. Magnetic resonance
  in medicine  \textbf{65}(2),  575--587 (2011)

\bibitem{yaman2020self}
Yaman, B., Hosseini, S.A.H., Moeller, S., Ellermann, J., U{\u{g}}urbil, K.,
  Ak{\c{c}}akaya, M.: Self-supervised learning of physics-guided reconstruction
  neural networks without fully sampled reference data. Magnetic resonance in
  medicine  \textbf{84}(6),  3172--3191 (2020)

\end{thebibliography}

\end{document}